\documentclass[12pt]{article}
\usepackage{amsmath,epsf,amssymb,latexsym,cite}
\newtheorem{theorem}{Theorem}
\newtheorem{prop}{Proposition}

\newtheorem{conjecture}{Conjecture}

\newif\iffigs\figstrue

\DeclareFontFamily{U}{rsf}{}
\DeclareFontShape{U}{rsf}{m}{n}{
  <5> <6> rsfs5 <7> <8> <9> rsfs7 <10-> rsfs10}{}
\DeclareMathAlphabet\Scr{U}{rsf}{m}{n}

\def\pplogo{\vbox{\kern-\headheight\kern -29pt
\halign{##&##\hfil\cr&{
\ppnumber}\cr\rule{0pt}{2.5ex}&\ppdate\cr}
}}
\makeatletter
\def\ps@firstpage{\ps@empty \def\@oddhead{\hss\pplogo}%
  \let\@evenhead\@oddhead 
}
\def\maketitle{\par
 \begingroup
 \def\thefootnote{\fnsymbol{footnote}}
 \def\@makefnmark{\hbox{$^{\@thefnmark}$\hss}}
 \if@twocolumn
 \twocolumn[\@maketitle]
 \else \newpage
 \global\@topnum\z@ \@maketitle \fi\thispagestyle{firstpage}\@thanks
 \endgroup
 \setcounter{footnote}{0}
 \let\maketitle\relax
 \let\@maketitle\relax
 \gdef\@thanks{}\gdef\@author{}\gdef\@title{}\let\thanks\relax}
\makeatother

\def\C{{\mathbb C}}

\def\Z{{\mathbb Z}}

\def\Hom{\operatorname{Hom}}

\def\End{\operatorname{End}}
\def\Ind{\operatorname{Ind}}

\def\tr{\operatorname{tr}}

\def\Lie{\operatorname{Lie}}

\def\Sl{\operatorname{SL}}
\def\Gl{\operatorname{GL}}

\def\SU{\operatorname{SU}}
\def\GU{\operatorname{U{}}}

\def\diag{\operatorname{diag}}

\def\CY{Calabi--Yau}

\def\cM{{\Scr M}}

\def\cG{{\Scr G}}

\def\labto#1{\mathrel{\mathop\to^{#1}}}

\begin{document}
\setcounter{page}0
\def\ppnumber{\vbox{\baselineskip14pt\hbox{DUKE-CGTP-00-18}
\hbox{hep-th/0009042}}}
\def\ppdate{September 2000} \date{}

\title{\LARGE D-branes, Discrete Torsion\\
and the McKay Correspondence\\[10mm]}
\author{
Paul S. Aspinwall and M. Ronen Plesser\\[10mm]
\normalsize Center for Geometry and Theoretical Physics, \\
\normalsize Box 90318, \\
\normalsize Duke University, \\
\normalsize Durham, NC 27708-0318\\[10mm]
}

{\hfuzz=10cm\maketitle}

\def\Large{\large}
\def\LARGE{\large\bf}


\begin{abstract}
We analyze the D-branes of a type IIB string theory on an orbifold
singularity including the possibility of discrete torsion following
the work of Douglas et al. First we prove some general results about
the moduli space of a point associated to the ``regular
representation'' of the orbifold group. This includes some analysis of
the ``wrapped branes'' which necessarily appear when the orbifold
singularity is not isolated. Next we analyze the stringy
homology of the orbifold using the McKay correspondence and the
relationship between K-theory and homology. We find that discrete
torsion and torsion in this stringy homology are closely-related
concepts but that they differ in general.
Lastly we question to what extent the D-1 brane may be thought of as
being dual to a string.
\end{abstract}

\vfil\break


\section{Introduction}    \label{s:int}

D-branes as probes of string theory in nontrivial backgrounds have
been a source of useful insights since \cite{BDS:prob}.  In 
\cite{DM:qiv} Douglas and Moore used a quotient construction
to study the dynamics of branes at an orbifold singularity.  The
construction involves a choice of a representation of the quotient
group on the Chan--Paton indices.
This leads to a supersymmetric gauge theory on the worldvolume (as we
review later on).  The moduli space of vacua of this theory contains an
approximation to the geometry of the transverse space in which the
brane is free to move.  The parameters in the theory are determined by
the closed-string background; the parameter space thus probes the
moduli space of closed-string vacua.
We refer to \cite{Doug:Dgeom} and references therein for a
larger account of this work. 

Let us consider a type IIB string on the local form of an orbifold
$\C^n/\Gamma$ for some finite group $\Gamma\subset\SU(n)$.  It has
been clear from the early days of this subject that D-branes give a
particularly direct physical picture of the ``McKay
Correspondence''. That is, there is a relationship between the
even-dimensional homology, or more properly K-theory, of the
resolution of such an orbifold and the representation theory of
$\Gamma$. (We refer to \cite{Reid:McK} for a review of the mathematics
of the McKay correspondence.) It was realized in \cite{DGM:Dorb} that
by computing the masses of wrapped branes as one blew-up a
singularity, one could relate the homology of the blow-up to
representations of $\Gamma$. This correspondence was made much more
explicit in \cite{DG:fracM} where the explicit map between homology
and representation theory was computed.

Douglas \cite{D:disctor} was able to extend the work of \cite{DM:qiv}
by considering {\em projective\/} representations. It is very natural
to associate this degree of freedom with Vafa's discrete torsion
\cite{Vafa:tor}. It would be nice to extend the McKay correspondence to
include discrete torsion. The work of Gomis \cite{Gomis:dt} is related
to this question and we explore the subject further in this
paper. There has also been some work associating discrete torsion with
monodromy acting on partial resolutions
\cite{Joyce:desing,BL:,Ruan:dt}. We will not discuss this here.

Our aim in this paper is to cover the following topics:
\begin{enumerate}
\item We will present a computation of the moduli space of vacua
determined by the ``regular representation'' and its relation to the
target space geometry in the general case.  This involves
extending some proofs by Sardo Infirri \cite{SI:I,SI:II} to
the case of discrete torsion. We will also discuss how wrapped branes
appear when the singularity is not isolated. Much has already been
said about these topics in the context of specific examples and
various generalizations. See, for
example, \cite{DDG:wrap,DG:fracM,DF:dt2,BJL:dtdef}. 
\item We will use the McKay correspondence for K-theory to {\em
define\/} a
theory of stringy homology on the orbifold itself. In the absence of
discrete torsion this agrees with the known results obtained by
studying the masses of wrapped branes.  The stringy
homology will exhibit torsion when, and only when, discrete torsion is
switched on. This construction will be closely related to observations in
\cite{Gomis:dt}. 
\item We will discuss if there is such a thing as an S-duality of the
type IIB string which literally exchanges the string and the D1-brane.
\end{enumerate}

Of particular note in the second case is the relationship between
torsion in homology and discrete torsion. This has been analyzed in
some cases (see, for example, \cite{AMG:stab}) but the general picture
remains unclear. At least for the local analysis of an orbifold we
hope to shed some light on this question. We will see that the torsion
in homology naturally contains a subgroup of $\Z_p$, where $p$
is the order of the element of discrete torsion within the group
$H^2(\Gamma,\GU(1))$. 

In section \ref{s:quiv} we set up many aspects of the group theory we
use and describe how to construct quiver diagrams. In section
\ref{s:mod} we give a general proof that the moduli space of the
D-brane associated with a point really is the target space
$\C^n/\Gamma$ only in the case of an isolated singularity. We give the
general description of the non-isolated case in terms of the familiar
``wrapped branes'' which are associated to induced representations of
$\Gamma$.

In section \ref{s:McKay} we set up the McKay correspondence for
orbifolds with discrete torsion. This amounts to defining what we mean
by homology.

The previous sections are clarified by reviewing some examples in
section \ref{s:eg}. We draw the quiver corresponding to the field
theory, analyze the possibilities for wrapped branes
and compute some stringy homology groups.

Finally in section \ref{s:conc} we emphasize the relationship between
torsion in homology and discrete torsion. We also discuss the
S-duality of the type IIB string. 


\section{The $\Gamma$-Equivariant D-brane Quiver}  \label{s:quiv}

In this section we will encode the data associated to a D-brane on an
orbifold in terms of a quiver diagram.  We will be general here and
present the analysis for any representation of the orbifold group and
any choice of discrete torsion.  For definiteness, we will discuss the
case mentioned above, of D3-branes near a singularity of the local
form $\C^3/\Gamma$.  The worldvolume theory is then an ${\cal N}=1$
supersymmetric theory in four dimensions.
Our analysis follows the discussion 
in \cite{SI:II} with the (small) modifications required to incorporate
discrete torsion.

The orbifold theory is constructed following \cite{DM:qiv} as a
quotient from a theory of branes on the covering space.  Let the brane
be located at a point in $\C^n$ which is fixed under all of
$\Gamma$. We will assume this point is the origin. Let $R$ be an
$m$-dimensional representation of $\Gamma$.  We
then consider $m$ branes on the covering space.  The low--energy
dynamics on the brane worldvolume is thus a $\GU(m)$ gauge
theory\footnote{The diagonal $\GU(1)$ always decouples in the gauge
theory so the effective gauge group is only really $\SU(m)$.} with
${\cal N}=4$ supersymmetry in four dimensions.  In terms of the ${\cal
N}=1$ supersymmetry that is unbroken by the quotient, this is a gauge
theory with three chiral multiplets $X_i$ in the adjoint representation
(corresponding to the three complex transverse coordinates) and a
superpotential given by 
\begin{equation}
	W = \tr\left(X_1[X_2,X_3]\right).	\label{eq:Wcov}
\end{equation}
The representation $R$ defines a {\em lift\/} of the action of
the orbifolding group on $\C^n$ to the gauge group.  The orbifold
theory is obtained \cite{DM:qiv} by projecting onto the invariant
degrees of freedom under the combined action of $\Gamma$. The theory
we obtain under the quotienting process depends very much on $R$.
To begin with, and to fix our notations, we take $R$ to be a
linear\footnote{Non-projective.}
representation, corresponding to the absence of discrete torsion.

The quotient theory will also be a gauge theory.  Because the gauge
fields live in the directions tangent to the brane, $\Gamma$ acts on
these only via $R$.  The gauge fields in the quotient theories will
generate the subgroup of $\cG=\GU(m)$ which is invariant under this
action
\begin{equation}				\label{eq:Inva}
  R(\gamma)A R(\gamma)^{-1}=A,
\end{equation}
for all $\gamma\in\Gamma$.  To facilitate the following discussions we will
introduce some fancier notation.  First let us complexify the
gauge group. Our initial
complexified gauge group will be $\cG_{\C}=\Gl(m,\C)=\End(\C^m)$. The
invariant complexified gauge group $\cG_{\C}^\Gamma$ is then generated by
all matrices $A$ which satisfy (\ref{eq:Inva}).  We write this as 
\begin{equation}
  \cG_{\C}^\Gamma = (\End R)^\Gamma.
\end{equation}
These gauge fields will appear in appropriate supermultiplets.  In our
case we find vector multiplets.

Additional degrees of freedom in the low-energy theory come from the
invariant degrees of freedom contained in the chiral multiplets.
$\Gamma$ acts on these, in addition to its
embedding in $\cG$, via the $n$-dimensional representation $Q$
determining its action on $\C^n$.  The invariant degrees of freedom
now satisfy 
\begin{equation}
   \sum_{j=1}^nQ_{ij}(\gamma)R(\gamma)X_jR(\gamma)^{-1}=X_i,
\end{equation}
for all $\gamma\in\Gamma$.  In our more abstract notation this is
written as 
\begin{equation}
  X\in(Q\otimes\End R)^{\Gamma}.
\end{equation}
In addition to the gauge couplings, these interact via a cubic
superpotential that is simply the restriction to the invariant fields
of the superpotential (\ref{eq:Wcov}).

To allow for discrete torsion we allow $R$ to be a {\em projective\/}
representation. That is
\begin{equation}
  R(\gamma)R(\mu)=\alpha(\gamma,\mu)R(\gamma\mu),	\label{eq:Proj}
\end{equation}
where $\alpha(\gamma,\mu)\in\GU(1)$.\footnote{This implies that
$\alpha(1,\gamma)=\alpha(\gamma,1)=1$ for any $\gamma$.}  The
associativity of group multiplication is consistent with this provided 
\begin{equation}
	\alpha(\gamma,\mu)\alpha(\gamma\mu,\rho) =
	\alpha(\gamma,\mu\rho)\alpha(\mu,\rho),
\end{equation}
which is precisely the condition that 
$\alpha(\gamma,\mu)$ is a group 2-cocycle. Further, it is clear that 
two cocycles related by 
\begin{equation}
	\alpha'(\gamma,\mu) =
	{\beta(\gamma)\beta(\mu)\over\beta(\gamma\mu)}
	\alpha(\gamma,\mu)
\end{equation}
for any map $\beta:\Gamma\to \GU(1)$ lead to equivalent conditions
(\ref{eq:Proj}).  This shows that the allowed representations are
determined by $\alpha$ as an element of $H^2(\Gamma,\GU(1))$.
A choice of a cohomology class $\alpha$ then determines the phases
$\epsilon(\gamma,\mu)$ associated to twisted sectors in Vafa's
formulation \cite{Vafa:tor} by 
$\epsilon(\gamma,\mu)=\alpha(\gamma,\mu)\alpha(\mu,\gamma)^{-1}$ (note
this need only be defined for commuting $\gamma,\,\mu$). We
refer to \cite{D:disctor,Gomis:dt} for more details.  

We can now repeat the calculation of the quotient theory as above.  We
find it useful to resort to fairly
algebraic language. Any standard text on the representation theory of
finite groups should explain all the terms we use.

Let us introduce the $\C$-algebra $\C^\alpha\Gamma$ defined as
follows. Firstly we let $e_\gamma$, $\gamma\in\Gamma$, be a basis for
$\C^\alpha\Gamma$. That is, any element may be written uniquely as a
sum
\begin{equation}
  \sum_{\gamma\in\Gamma} c_\gamma e_\gamma,
\end{equation}
where $c_\gamma\in\C$. We define a distributive product by 
\begin{equation}
  e_\gamma e_\mu = \alpha(\gamma,\mu)e_{\gamma\mu}.
\end{equation}
The standard theory of representations and modules now says that {\em
any projective representation of $\Gamma$ can be written as a
$\C^\alpha\Gamma$-module.} Naturally we may obtain
linear
representations by using the trivial $\alpha$ to obtain
$\C\Gamma$-modules. 

Let $R_l$ represent the {\em irreducible\/} representations of
$\Gamma$ twisted by $\alpha\in H^2(\Gamma,\GU(1))$.  It is well-known
that irreducible {\em linear\/} representations of $\Gamma$ are
counted by the number of conjugacy classes of $\Gamma$. There is a
similar concept for projective representations.  An
``$\alpha$-regular'' conjugacy class contains elements $\gamma$
satisfying $\alpha(\gamma,\mu) = \alpha(\mu,\gamma)$ for all
$\mu\in\Gamma$ such that $\gamma\mu=\mu\gamma$. The number of
irreducible projective representations is equal to the number of
$\alpha$-regular conjugacy classes.

We may
view $R_l$ as {\em simple\/} $\C^\alpha\Gamma$-modules. We refer to
\cite{Karpil:proj} for an exposition of the properties of irreducible
projective representations. We may then decompose
\begin{equation}
  R = \bigoplus_{l}V_l\otimes R_l,
\end{equation}
where $V_l$ is a linear vector space whose dimension is determined by
the number of times a given irreducible representation appears in $R$.

Let us use $\Hom_\Gamma$ to denote the group of homomorphisms of
$\C^\alpha\Gamma$-modules. We may now deduce that for $A\in\cG_{\C}^\Gamma$
\begin{equation}
\begin{split}
  A &\in (\End R)^\Gamma\\
    &=\Hom_\Gamma(R,R)\\
    &=\Hom_\Gamma\left(\bigoplus_l R_l\otimes V_l,
	\bigoplus_m R_m\otimes V_m\right)\\
    &=\bigoplus_l \Hom(V_l,V_l)\\
    &=\bigoplus_l \End(V_l),
\end{split}
\end{equation}
where the penultimate step was obtained by using Schur's Lemma. Thus
if we denote $v_l=\dim(V_l)$ we see that
$\cG_{\C}^\Gamma=\prod_l\Gl(v_l)$. In other words, restricting back to the
compact real form, we have a gauge group
\begin{equation}
  \cG^\Gamma=\prod_l\GU(v_l).
\end{equation}

Since $Q$ is a linear representation of $\Gamma$ and $R$ is
a projective representation twisted by $\alpha$, it is not hard to
show that $Q\otimes R$ is a projective representation twisted by
$\alpha$. We may thus decompose
\begin{equation}
  Q\otimes R_l = \bigoplus_m A_{lm}\otimes R_m,
\end{equation}
where $A_{lm}$ are vector spaces.

We may then deduce
\begin{equation}
\begin{split}
  X &\in (Q\otimes\End R)^{\Gamma}\\
    &= \Hom_\Gamma(R,Q\otimes R)\\
    &= \Hom_\Gamma\left(\bigoplus_l R_l\otimes
      V_l, Q\otimes\left(\bigoplus_m R_m\otimes
      V_m\right)\right)\\
    &= \Hom_\Gamma\left(
      \bigoplus_l R_l\otimes V_l,
      \bigoplus_{mn}A_{mn}\otimes R_n\otimes V_m\right),\\
    &= \bigoplus_{lm} A_{lm}\otimes\Hom(V_l,V_m).
\end{split}
\end{equation}
As above, in addition to the gauge couplings, the chiral multiplets
interact via a cubic superpotential descended from (\ref{eq:Wcov}).  

A useful way of visualizing the structure of the resulting theory is
in terms of a {\em quiver}.
\begin{enumerate}
   \item There is a node for every irreducible representation
   $R_l$. Each node is labelled by $v_l=\dim V_l$. Each node
   represents a factor of $\GU(v_l)$ in the 
   gauge group $\cG^\Gamma$.
   \item From node $l$ to node $m$ we draw $a_{lm}=\dim A_{lm}$
   arrows. Each arrow represents a contribution of $\Hom(V_l,V_m)$
   to the space in which $X$ lives. That is, the arrow represents a
   $(\overline{\mathbf{v}}_l,\mathbf{v}_m)$ representation of 
   $\GU(v_l)\times\GU(v_m)$.
\end{enumerate}
Note that the numbers $a_{lm}$, and hence the arrows, are fixed by
$\Gamma$ and $\alpha$ and do not depend on $R$. The numbers $v_l$ do
depend on $R$. If $v_l=0$ for a particular node then one can
ignore that vertex and the associated arrows.
We will give several examples of quivers below.


\section{D-brane Moduli Space}     \label{s:mod}

One object of immediate interest is the moduli space of classical
vacua of the gauge theory.  The is the zero string-coupling limit of
the moduli space of the associated brane in the orbifold background.
The classical vacua are parameterized by values for the scalars in the
chiral multiplets, modulo gauge equivalence, solving $F$-term
equations (critical points of the superpotential) as well as the
$D$-term equations (zero moment maps for the action of $\cG^\Gamma$).
The details of these depend upon $R$, but since the quotient theory is
obtained from the covering theory simply be reducing the gauge group
to a subgroup of $\cG$ and setting some of the chiral fields to zero,
we can give a uniform description directly in terms of the covering
theory.  From (\ref{eq:Wcov}) we have the $F$-terms
\begin{equation}
[X_i,X_j]=0,\quad\text{for all $i,j$}   \label{eq:Fterm}
\end{equation}
where in a particular quotient we keep only the invariant components of
$X$.  The moment map for $\cG$ is simply
\begin{equation}
\sum_{i=1}^n[X^\dagger_i,X_i]=0.        \label{eq:Dterm}
\end{equation}
In the quotient theory, only the parts of this corresponding to
$\cG^\Gamma$ will be nontrivial.
One then divides the solution set of these equations by the gauge
group $\cG^\Gamma$ to obtain $\cM_R$. The key idea of the
physics of D-branes on the orbifold is the following
\begin{prop}
  The space $\cM_R$ represents the space of allowed positions of the
  D-brane associated to $R$ in the target space $\C^n/\Gamma$.
\end{prop}

The simplest case of using an irreducible representation for $R$
tends to give $\cM_R$ equal to a point --- the wrapped D-brane is
stuck at the origin. One obtains more interesting results by using
bigger representations.

\subsection{The regular representation}

Of particular note is the case when $R$ is given by putting
$R=\C^\alpha\Gamma$. (That is, we view $R=\C^\alpha\Gamma$ itself as a
$\C^\alpha\Gamma$-module.) When $\alpha$ is trivial, this is equivalent to
making $R$ the {\em regular representation}.\footnote{One may extend
the use of language to call $\C^\alpha\Gamma$ the regular
representation even when $\alpha$ is nontrivial. This was implicit in
\cite{D:disctor}.} Note that now \cite{Karpil:proj}
\begin{equation}
v_l=\dim(R_l).
\end{equation}
Taking the dimension of $\C^\alpha\Gamma$ as a vector space over $\C$,
this implies a result which will be useful later on:
\begin{equation}
  |\Gamma| = \sum_l v_l^2.   \label{eq:dim}
\end{equation}

We may now prove the following theorems:
\begin{theorem}   \label{th:a}
  If $R=\C^\alpha\Gamma$ and $\Gamma$ acts freely on $\C^n$
  outside the origin then $\cM_R$ is
  the orbifold $\C^n/\Gamma$ itself.
\end{theorem}
\begin{theorem}   \label{th:b}
  If $R=\C^\alpha\Gamma$ and $\Gamma$ does not act freely on $\C^n$
  outside the origin then $\cM_R$ consists of more than one
  component, one of which is the orbifold $\C^n/\Gamma$ itself.
\end{theorem}

To do this we follow the proof of theorem 4.2 in \cite{SI:I} with a
little modification to allow for discrete torsion.

\def\adA{\mathsf{A}}
First let $\adA_i$ denote the operator $\mathrm{ad}(X_i)$, that
is $\adA_i(X)=[X_i,X]$. We also have the Hermitian conjugate 
operator $\adA^\dagger_i$. The equations (\ref{eq:Fterm}) and
(\ref{eq:Dterm}) together with the Jacobi identity show that
\begin{equation}
  \sum_i \adA^\dagger_i\adA_i(X^\dagger_j)=0.
\end{equation}
It is a basic fact of linear algebra that $\adA^\dagger_i\adA_i$ is a
positive operator. Therefore the above implies that
$\adA^\dagger_i\adA_i(X^\dagger_j)=0$ for any choice of $i$ or
$j$. That is, $\adA_i(X^\dagger_j)=0$ for any choice of $i$ or
$j$. In particular
\begin{equation}
  \adA_i(X^\dagger_i)=[X^\dagger_i,X_i]=0.
\end{equation}
This means that each matrix $X_i$ is {\em normal\/} and can be
diagonalized by conjugation by a unitary matrix $U \in
\cG=\GU(m)$. Furthermore (\ref{eq:Fterm}) implies that $X_i$ can be
simultaneously diagonalized for all $i$.  

Let $v_1\in R$ be a simultaneous eigenvector of $X_i$ with eigenvalues
$\lambda_1^{(i)}$. Denote the basis of $Q=\C^n$ by $q_i$,
$i=1,\ldots,n$. We then write
\begin{equation}
  \lambda_1 = \sum_i \lambda_1^{(i)}q_i\in Q.
\end{equation}

Now, the fact that $X\in(Q\otimes\End R)^{\Gamma}$ tells us that
\begin{equation}
  \sum_j Q_{ij}(\gamma)R(\gamma)X_j = X_iR(\gamma)
\end{equation}
and so
\begin{equation}
\begin{split}
  \sum_j Q_{ij}(\gamma)R(\gamma)X_jv_1 &= X_iR(\gamma)v_1\\
    &= \sum_j Q_{ij}(\gamma)\lambda_1^{(j)}R(\gamma)v_1.
\end{split}
\end{equation}
In other words $R(\gamma)v_1$ is an eigenvector of $X$ with eigenvalue
$Q(\gamma)\lambda_1$. {\bf Note in particular that the eigenvalues must
always appear as $\Gamma$-orbits in $Q$.}
Let us denote $R(\gamma)v_1$ by $v_\gamma$.

Now let us first assume that the action of $\Gamma$ on $Q$ has no
fixed points away from the origin. Then assuming $v_1$ is non-zero,
the eigenvalues are all distinct and so
the vectors $v_\gamma$ will form an orthogonal basis for
$R=\C^\alpha\Gamma$. We may fix this basis to be orthonormal.

Let us choose a fixed orthonormal basis for $\C^\alpha\Gamma$ by
fixing $e_1$ and 
then defining $e_\gamma=\gamma e_1$. The unitary matrix $U$
diagonalizing the $X_i$ can be taken 
to rotate the $v_\gamma$ basis into the $e_\gamma$
basis. That is, $Uv_\gamma=e_\gamma$, for all $\gamma$. Note that
for any $\mu\in\Gamma$,
\begin{equation}
\begin{split}
  R(\mu)Uv_\gamma &= R(\mu)e_\gamma\\
	&= \mu e_\gamma\\
        &= \alpha(\mu,\gamma) e_{\mu\gamma},
\end{split}
\end{equation}
and
\begin{equation}
\begin{split}
  UR(\mu)v_\gamma &= UR(\mu)R(\gamma)v_1\\
	&= \alpha(\mu,\gamma)UR(\mu\gamma)v_1\\
	&= \alpha(\mu,\gamma)Uv_{\mu\gamma}\\
        &= \alpha(\mu,\gamma)e_{\mu\gamma}.
\end{split}
\end{equation}
Thus $U$ and $R(\mu)$ commute which implies that $U\in\cG^\Gamma$.

So now, we see that a point $\cM_R$ is determined by the eigenvalues
$\lambda$, up to permutations.  What we have shown above is that this
set is precisely in one-to-one correspondence with $\Gamma$-orbits in
$Q$.  That the correspondence is onto is easy to see by taking
diagonal matrices.  This proves theorem~\ref{th:a}.

We now come to the more interesting case of a non-isolated fixed point
at the origin. Again the analysis of Sardo Infirri \cite{SI:I} applies
even with discrete torsion switched on.

A fairly simple argument in linear algebra may be used to show that
$Q/\Gamma\subset\cM_R$. Let $x$ be any point in $Q$ and let $\Gamma_x$
be the subgroup of $\Gamma$ which fixes $x$. We assume $\Gamma_x$ is
trivial for a generic $x\in Q$. Given $x$, we may construct $X$ such
that its eigenvalues are given by the $\Gamma$-orbit of $x$ with each
eigenvalue appearing with multiplicity $|\Gamma_x|$. Now this matrix
will have eigenvectors forming an 
orthonormal basis of $R$ and we again recover the above
construction. Note also that if $x$ lies in an open neighbourhood such
that $\Gamma$ acts freely on this neighbourhood then this inclusion
$Q/\Gamma\hookrightarrow\cM_R$ is locally a homeomorphism on this
subset. Thus $Q/\Gamma$ appears as a component of $\cM_R$.

A key point however is that if $\Gamma_x$ is nontrivial we may choose
$X$ to have eigenvalues $x$ of multiplicity {\em less than\/} $|\Gamma_x|$.
Suppose we have a point $x\in Q$ away from the origin which
is fixed by $\Gamma_x\subsetneq\Gamma$. The $\Gamma$-orbit of $x$ will
now have fewer than $|\Gamma|$ points. We may build an $X$ which is
diagonal and whose eigenvalues correspond to the $\Gamma$-orbit of $x$
together with zero for the remaining eigenvalues. This is clearly a
valid $X$ which lies outside any one-to-one correspondence between
$\cM_R$ and $Q/\Gamma$. Thus we see that theorem~\ref{th:a} {\em must\/} fail
if the quotient singularity is not isolated at the origin.
This proves theorem~\ref{th:b}.

\subsection{Mobile wrapped branes}

These extra branches of $\cM_R$ correspond to the wrapped
branes discussed in \cite{DDG:wrap}. We have shown that one must
always obtain extra branches of the moduli space corresponding
to wrapped branes if the quotient singularity in $Q/\Gamma$ is not isolated.

The analysis of \cite{SI:I} tells us exactly how to describe these
extra branches. Again the description
remains valid with discrete torsion switched on. Suppose $x$ is a
point away from the origin fixed
by $\Gamma_x$ as above. We may
consider an $\alpha$-twisted projective representation $R_x$ of
$\Gamma_x$. This ``induces'' a representation of $\Gamma$ of the form
\begin{equation}
  \Ind_{\Gamma_x}^\Gamma R_x =
  \C^\alpha\Gamma\otimes_{\C^\alpha\Gamma_x} R_x,
\end{equation}
(where $\C^\alpha\Gamma$ is viewed as a left $\C^\alpha\Gamma$-module
and a right $\C^\alpha\Gamma_x$-module).
This induced representation will represent a ``wrapped brane'' which
can be constrained to live along the fixed locus of $\Gamma_x$ if
$R_x$ is chosen to have sufficiently low dimension. Using an irreducible
representation for $R_x$ will always yield such an example. If we take
the ``regular representation'' $\C^\alpha\Gamma_x$ for $R_x$ then clearly
\begin{equation}
\begin{split}
  \Ind_{\Gamma_x}^\Gamma R_x &=
  \C^\alpha\Gamma\otimes_{\C^\alpha\Gamma_x} \C^\alpha\Gamma_x\\
  &= \C^\alpha\Gamma.
\end{split}
\end{equation}
This shows how a suitable sum of these wrapped branes ``stuck'' along
the fixed point set will combine to give the brane associated to
$\C^\alpha\Gamma$.  This represents the
set of wrapped branes corresponding to one or more of the extra
branches of the moduli space. Looking at all fixed subspaces in the
orbifold will account for all the branches.

Conversely, given a representation $R$ of $\Gamma$, we may describe
the moduli space associated to $R$ as follows.  $R$ can be induced by
various subgroups.  We need to consider each {\em minimal\/} such
subgroup, in the sense that it contains no other subgroups that induce
$R$.  The moduli space corresponding to $R$ will be the union of the
fixed point sets of these minimal subgroups.

We will give several examples of these induced representations in section
\ref{s:eg}.


\section{The McKay Correspondence}   \label{s:McKay}

\subsection{An attempt with homology}

The general idea of the McKay correspondence is as follows. Let
$Y\to\C^n/\Gamma$ be a crepent (i.e., maintaining the \CY\ condition)
maximal blow-up. The homology of $Y$ is then closely related to the
representation theory of $\Gamma$.

One of the most straight-forward but crude ways of seeing this
correspondence in D-branes arises from the work of
\cite{DM:qiv,DGM:Dorb}. Let us begin by considering the particular
case of the representation $R=\C^\alpha\Gamma$ as in the previous section.
Each node in the quiver associated to a D-brane on a orbifold can be
associated to a potential Fayet-Iliopoulos term in the D-brane action. It is
well-established (see \cite{SI:I,SI:II,DGM:Dorb,GLR:nonab} for
example) that adding such terms into the action can result in a blow-up
of $\cM_R=\C^n/\Gamma$. Note that this fact remains true even when discrete
torsion is switched on.\footnote{The choice of discrete torsion
considered in most 
examples in the literature results in a system so tightly constrained
that no blow-ups remain. A sufficiently general example will still
have blow-ups however.}
The way to picture these blow-ups is most easily seen in terms of {\em
moment maps}. Note 
that
\begin{equation}
   \rho=\sum_{i=1}^n[X^\dagger_i,X_i],
\end{equation}
represents a moment map $\rho:X\to\Lie^*\cG^\Gamma$. The effect of
adding Fayet-Iliopoulos terms to the action is to effectively shift
this moment map $\rho\to\rho-\zeta$ for some new moment map $\zeta$
lying in the centre of $\Lie^*\cG^\Gamma$. We refer to \cite{SI:I} for
more details. The centre of $\cG^\Gamma$ corresponds to $\GU(1)$'s
associated to each vertex of the quiver. We have fixed our basis of
representations $R_l$ to label each of these vertices. Thus we may expand
\begin{equation}
  \zeta = \sum_{p}\zeta_p R^*_p,
\end{equation}
where $R^*_p$ is a basis {\em dual\/} to the representations $R_l$.

One may add the same terms to the action by adding insertions of
closed string twist 
fields. Let us review the case for no discrete torsion.
An appendix in \cite{DM:qiv} showed how these twist fields are
related precisely to the Fayet-Iliopoulos terms. Let $\phi_\gamma$ be a
field in the NS-NS sector of the string theory twisted by $\gamma$. Then 
$\zeta_p=\sum_\gamma\tr R^*_p(\gamma)\phi_\gamma$, where the sum is taken
over the conjugacy classes of $\Gamma$. Using the orthogonality of
the characters $\chi_p^*(\gamma)=\tr R^*_p(\gamma)$ we may rewrite
this as
\begin{equation}
  \phi_\gamma = \sum_p \chi_p(\gamma)\zeta_p,    \label{eq:M1}
\end{equation}
(where we have rescaled $\phi_\gamma$).

Including the possibility of discrete torsion actually makes little
difference to this argument. 
It was argued in \cite{Vafa:tor,VW:tor,Gomis:dt} that the twisted
strings in a theory with discrete torsion are precisely those
corresponding to $\alpha$-regular conjugacy classes. Now the theory of
characters of projective representations and $\alpha$-regular
conjugacy classes is essentially identical to the usual theory of
characters and conjugacy classes. We refer to \cite{Karpil:proj} for
more details. In fact, equation (\ref{eq:M1}) is still true with
discrete torsion switched on so long as we use the relevant notions of
projective representations.

For a given $\gamma\in\Gamma$, we may write the eigenvalues of
$\gamma$ as $\exp(2\pi ia_1)$, $\exp(2\pi ia_2)$, \ldots, $\exp(2\pi ia_n)$
where $0\leq a_i<1$. It is then a well-known result of topological
field theory that we may then associate $\phi_\gamma$ with an element
of $H^{2w(\gamma)}$ with 
\begin{equation}
  w(\gamma)=\sum_i a_i.   \label{eq:a}
\end{equation}
(Presumably this cohomology group
relates to cohomology with compact support as our target space is not
compact.) Notable fields are those obeying $w(\gamma)=1$. These
correspond to marginal deformations and relate to deformations of the
K\"ahler form in $H^2$. The formalism of topological field theory extends
this to the case $w(\gamma)\neq1$.

To return to the basis $R_l$ naturally associated to the vertices of
our quiver we need to take the dual of this mapping and so we relate
the group generated by the representations of $\Gamma$ to the {\em
homology\/} of the resolution. We may then state our first draft
of the D-brane McKay correspondence:
\begin{conjecture}
  Let $Y$ be a maximal crepent resolution of $\C^n/\Gamma$ allowed by
  a given choice (possibly trivial) of discrete torsion. For any
  $\gamma\in\Gamma$ we associate a cycle in $H_{2w(\gamma)}(Y)$ to the
  D-brane with representation $\sum_l\chi_l^*(\gamma)R_l$.
\end{conjecture}

One can have immediate success with this conjecture by considering the
case of the untwisted sector given by $\gamma=1$. This conjecture then
says that the associated homology class lies in $H_0$ and corresponds
to $\sum_l\dim(R_l)R_l=\C^\alpha\Gamma$. In section~\ref{s:mod} we saw
how the representation $\C^\alpha\Gamma$ had the entire target space
$\C^n/\Gamma$ as (at least one component) its moduli space. This
agrees beautifully with it being the moduli space of a single
point. This idea of associating a point with $\C^\alpha\Gamma$ indeed goes
back to the original work of \cite{DM:qiv}.

It is not hard to see from arguments along the lines of
section~\ref{s:mod} that taking $R$ to be two copies of $\C^\alpha\Gamma$
will give a symmetric product of $\C^n/\Gamma$ with itself (again with
potentially further branches which we ignore for now). This is thus
the moduli space of two points. Clearly this argument works for any
number of points. We will therefore take this conjecture to
be true in the case of $H_0$. That is, {\bf a point is always given by the
representation $\C^\alpha\Gamma$.}

As soon as one tries to go beyond this simple case one immediately
runs into difficulty. The reason for this is clear. In the language of
\cite{W:phase,AGM:II}, the twisted fields $\phi_\gamma$ are a natural
basis of cohomology in the {\em orbifold\/} phase whereas one would
normally like to picture the homology in question as living in the
large radius {\em \CY\/} phase. One therefore needs to change basis
between these phases to get the right statement for the McKay
correspondence. This change of basis was essentially described in
\cite{AGM:sd} and has been described in exactly this context in
\cite{DG:fracM}. We refer to these references for more details.

Having said all this, it might come as a surprise that the case of $H_0$
worked out so nicely. Why didn't we have to mix in other twist fields
to to get the classical $H_0$? A special r\^ole is given to $H_0$ 
essentially because we are working in a {\em noncompact\/} example of
$\C^n/\Gamma$. We will not attempt a general proof here but it can be
seen from the examples of \cite{AGM:sd,DG:fracM} that in this case one
has a simple constant as one possible solution of the Picard-Fuchs
equation. This is not the case for a compact \CY\ manifold and so one
assumes that in such a case $H_0$ would mix freely with the other
dimensions.	

Even after allowing for this mixing within the homology, our
conjecture is not really satisfactory.  In particular, considering
wrapped branes will not give us information on torsion classes in the
homology.  We need to discuss K-theory to improve matters.

\subsection{K-theory}  \label{ss:K}

One of the most general forms of the McKay correspondence was given in
\cite{BKM:MisM} which states the McKay correspondence in terms of
derived categories of coherent sheaves. This in turn implies a
statement about K-theory. Although it is clear that
derived categories should play an important r\^ole in string theory we
will restrict ourselves to the weaker language of K-theory in this
paper.

As is well-known, K-theory is similar to (singular) cohomology but can
differ. In the context of D-branes it is now well-established that
K-theory is the more relevant notion \cite{MM:K,W:K}. It is not
surprising therefore that D-branes give a nice picture of the McKay
correspondence when the language of K-theory is used. 

Consider a type IIB string theory on the orbifold $\C^n/\Gamma$ with a
choice of discrete torsion $\alpha$. Let $D_0$ denote the lattice of
resulting D-brane charges. Crudely stated, we know that $D_0$ is associated to
$H^{\textrm{even}}(\C^n/\Gamma,\Z)$ in some way.

Let us assert our version of the McKay correspondence:
\begin{prop}   \label{p:McKK}
  The lattice $D_0$ is isomorphic to the free abelian group generated by the
  $\alpha$-twisted irreducible projective representations of $\Gamma$.
\end{prop}
When $\alpha=1$ this reduces to the usual McKay correspondence and the
relationship between D-branes and K-theory. For nontrivial $\alpha$ it
is probably difficult to prove this proposition given the current
status of our definitions of string theory in a singular space. Indeed
one may wish to regard this proposition as a definition of $D_0$.

In order to get a better feeling for this conjecture we need to relate
things to our discussion of homology above. As we saw, it really is
homology rather than cohomology which is most naturally associated to
the representation theory of $\Gamma$. 
Because of this, let us introduce a group $K_0$ which is a homological
version of 
K-theory. It is usual to define K-theory as an abstract cohomology
theory in the Eilenberg-Steenrod sense but one may also define the
corresponding theory of abstract homology. 

One may proceed as follows. Define
$K_{\textrm{even}}(\mathrm{point})=\Z$ and
$K_{\textrm{odd}}(\mathrm{point})=0$. Now define $K_*$ as a theory of
homology in the Eilenberg-Steenrod sense (see, for example, section
2.3 of \cite{Hatch:top}). This is sufficient to define
$K_{\textrm{even}}$ and $K_{\textrm{odd}}$ for any topological
space. Replacing homology by cohomology here would give the usual
K-theory.

For a fairly large class of toplogical spaces one can relate $K_0$ to
the usual {\em singular\/} homology $H_*$. This may be done by using
the Atiyah-Hirzebruch spectral sequence \cite{AH:K} rewritten in terms
of homology. This has also been used recently in the physics
literature in \cite{DMK:KMbig}. We simply quote the following

\begin{theorem}
Let $Y$ be a finite simplicial complex. There exists a homology
spectral sequence with
\begin{equation}
   E^2_{p,q} = H_p(Y,K_q(\mathrm{point})),
\end{equation}
which converges to give $E^\infty_{p,q}\cong K_{p+q}(Y)$.
\end{theorem}
This theorem encodes two ways in which $K_0(Y)$ may differ from
$H_{\textrm{even}}(Y,\Z)$. Firstly one may have nontrivial
differentials $\partial_r:E^r_{p,q}\to E^r_{p-r,q+r-1}$. The effect of
this is to kill elements of $K_0(Y)$ which appear in
$H_{\textrm{even}}(Y,\Z)$. While this certainly can happen, such
effects tend not to occur until one considers homology of a large
dimension. We are mainly concerned with 0-cycles and 2-cycles in this
paper and {\em we will ignore such a possibility.}\footnote{For the
dual cohomology spectral sequence one may analyze the differentials in
terms of ``cohomology operations''. This severely restricts the
allowed maps. Indeed, proposition 4.82 of \cite{Hatch:top} may be used
to rule out the possibility of differentials in the spectral sequence
from $H^0$ or $H^2$.}  We thus assume that $E^2_{p,q}=E^\infty_{p,q}$.

Of more interest is the fact that $E^\infty_{p,q}$ is associated to a
filtration. Ignoring the effect of the differentials, there is a sequence of
inclusions\footnote{Note this this filtration is in the opposite
direction to the usual K-theory associated with cohomology.}
\begin{equation}
  K_0 =K_0^\infty\supset\ldots\supset K^4_0\supset K^2_0\supset K^0_0,
\end{equation}
where
\begin{equation}
  H_n \cong K^n_0/K^{n-2}_0,    \label{eq:grad}
\end{equation}
and $K^{-2}_0=0$. 

Returning to the world of string theory, we know that singular
(co)homology is {\em not\/} the relevant notion for describing string
states on an orbifold with, or without, torsion. One needs to use
``stringy (co)homology''. We will assert that the group $D_0$
essentially plays the r\^ole an object like $K_0$ except that it is
related to stringy homology via the Atiyah-Hirzebruch spectral
sequence. {\em From now on we will use the symbol $K_0$ to refer to this
stringy object.} Note that $K_0$ knows about discrete torsion.
This is similar to the approach of \cite{Gomis:dt}. Preliminary
aspects of the mathematics of this object have been conjectured
\cite{Cald:dK}. We use equation (\ref{eq:grad}) as our {\em definition\/} of
stringy homology.

Note that (\ref{eq:grad}) is not quite the same thing as saying
\begin{equation}
  K_0 = \bigoplus_{i \mathrm{\ even}} H_i,   \label{eq:wrong}
\end{equation}
as we will see shortly. Indeed
(\ref{eq:wrong}) is false in general. Having said that,
(\ref{eq:wrong}) is correct if one considers {\em rational\/} (or real or
complex) coefficients rather than integers. In this way we really
should see a McKay correspondence for homology from D-branes if we ignore
considerations such as torsion.

Finally in this section we would like to prove the following:
\begin{theorem}   \label{th:dt}
The stringy homology of $\C^n/\Gamma$ contains a torsion group
of order $p$ as a subgroup, where $p$ is the order of the discrete torsion
$\alpha$ in the group $H^2(\Gamma,\GU(1))$. Furthermore, the group $H_2$ is
torsion-free if $\alpha$ is trivial.
\end{theorem}

Suppose first that $\alpha$ is trivial.
Note that $K_0$ and thus all the subgroups $K_0^n$ are
free. Next note that if $\alpha$ is trivial then one of the
representations of $\Gamma$ is the trivial representation, $R_0$,
which is one dimensional. Now the regular representation $\C\Gamma$
breaks up into a sum $\oplus_l\dim(R_l)R_l$ and so $R_0$ appears with
multiplicity one. Since $\C\Gamma$ generates $K_0^0$ we may construct
$H_2=K_0^2/K_0^0$ from $K_0^2$ simply by eliminating $R_0$ as a
generator. Thus $H_2$ is free and we prove the second part of the
theorem.

The proof of the first part of the theorem requires the introduction
of some more technicalities and the reader may wish to skip the rest
of this section. The ``representation group'', $\hat\Gamma$, of a
group $\Gamma$ over the field $\C$ is defined as follows. Suppose we
have the following central extension of $\Gamma$:
\begin{equation}
\renewcommand{\arraystretch}{0.1}
  1 \to A \labto{i} \hat\Gamma\begin{array}{c}{\scriptstyle s}\\ 
  \curvearrowleft\\{\scriptstyle j}\\\to\\ 
  \vphantom{l}\\ \vphantom{m}\end{array}\Gamma \to 1,
\end{equation}
where $A$ is a finite group isomorphic to $H^2(\Gamma,\GU(1))$. 
The map $s$ is a set-theoretic map acting as a right-inverse to $j$. Given
any linear representation of $\hat\Gamma$ we may use $s$ to define a
projective representation of $\Gamma$, The group $\hat\Gamma$ is said
to be a representation group for $\Gamma$ over $\C$ if  any projective
representation may be lifted to a linear representation of
$\hat\Gamma$. It is established (see 3.3 in 
\cite{Karpil:proj}) that $\hat\Gamma$ exists for any $\Gamma$.

Let $R$ be an $N$-dimensional irreducible projective representation of $\Gamma$
twisted by a specific $\alpha\in H^2(\Gamma,\GU(1))$. We lift this to
an $N$-dimensional irreducible linear representation $\hat R$ of $\hat\Gamma$. 
Since $i(A)$ is central, the representation in $\hat R$ of any element
of $i(A)$ must be proportional to the identity matrix. There is
natural isomorphism $\Hom(A,\GU(1))\to H^2(\Gamma,\GU(1))$ which
explicitly gives $\hat R(i(A))$ as a subgroup of $\GU(1)$ for a given choice of
$\alpha$. 

Consider the projective representation ring of $\Gamma$. If $R_1$ is an
$\alpha_1$-twisted projective representation of $\Gamma$ and 
$R_2$ is an $\alpha_1$-twisted projective representation then
$R_1\otimes R_2$ is an ($\alpha_1+\alpha_2$)-twisted projective
representation where we use the natural group structure of
$H^2(\Gamma,\GU(1))$. Thus if $\alpha$ is of order $p$ in
$H^2(\Gamma,\GU(1))$, then the elements of $\hat R(i(A))$ must be of order
$p$. Indeed it must be that $\hat R(i(A))\cong\Z_p$.

Now note that one cannot have a one-dimensional
truly projective representation. This is because the choice of $\alpha$ in
such a representation would necessarily be the coboundary of a one-cocycle.
Taking determinants of the matrices $\hat R$ above, gives a
one-dimensional representation which is an submodule of the tensor
product $R^{\otimes N}$. This implies that the $N$th power of any
element of $A$ must be trivial in the representation $\hat R(i(A))$. {\em
This implies that $N$ is a multiple of $p$.} 

The above argument shows that all the irreducible projective
representations of $\Gamma$ have dimension equal to a multiple of
some integer $p$ given by the order of $\alpha$. This implies that the
representation 
$\C^\alpha\Gamma=\oplus_l\dim(R_l)R_l$ is $p$-divisible. Since $K^0_0$
is generated by 
$\C^\alpha\Gamma$, this means that $K_0/K^0_0$ has an element of
$p$-torsion. This proves the first part of the theorem.

Note that we have not proven that this torsion subgroup lives entirely
in $H_2$. The form of the filtration could conceivably force it to be
spread over many homology groups.  It would also be interesting if we
could prove that $p$ is the greatest common divisor of the dimensions
of the projective representations of $\Gamma$. This would show that
the torsion in $K_0/K^0_0$ is precisely $\Z_p$. We will not attempt
this here.


\section{Examples}   \label{s:eg}

\begin{table}
\def\p{\phantom{-}}
$$\begin{array}{c|cccc}
\multicolumn{1}{c}{}&R_0&R_1&R_2&R_3\\
\cline{2-5}
1&1&\p1&\p1&\p1\\
a&1&-1&\p1&-1\\
b&1&\p1&-1&-1\\
ab&1&-1&-1&\p1
\end{array}$$
  \caption{The irreducible representations of $\Z_2\times\Z_2$.}
  \label{tab:11}
\end{table}

\subsection{$\Z_2\times\Z_2$}    \label{ss:Z2xZ2}

We first consider everyone's favorite example of discrete torsion.
Let $\Gamma$ be the group $\Z_2\times\Z_2$ acting on $\C^3$ generated
by $a=\diag(-1,-1,1)$ and $b=\diag(1,-1,-1)$. It is well-known that
$H^2(\Z_2\times\Z_2,\GU(1))\cong\Z_2$. Hence there are two choices ---
one may have discrete torsion switched on or off. This example gives
rise to a non-isolated singularity. There are three complex lines of
$\Z_2$-fixed points passing through the origin.

\subsubsection{No discrete torsion}

Let us first discuss the case without discrete torsion. Many aspects
of this have been analyzed in the context of D-branes in
\cite{BRG:Z2} for example. Since this is an abelian group, there are as many
representations as there are group elements --- i.e., four. 
Each representation is one-dimensional and is listed in table~\ref{tab:11}.
We show the quiver in figure~\ref{fig:q10}. Note that if we have an arrow
going from vertex $i$ to vertex $j$ and another arrow from vertex $j$
back to vertex $i$ then we combine them into a single line in this
diagram. The simplifies the appearance of the diagrams in this
paper.\footnote{Indeed all the arrows appearing in this paper have
this property. This is certainly not true in general.}

\iffigs
\begin{figure}
  \centerline{\epsfxsize=5cm\epsfbox{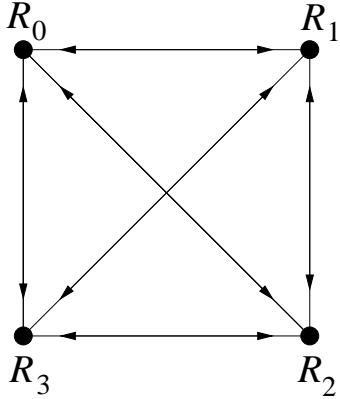}}
  \caption{The quiver for $\Z_2\times\Z_2$ with no discrete torsion.}
  \label{fig:q10}
\end{figure}
\fi

We see therefore that $K_0$ is the free abelian group generated by
four elements which we call $k_0,\ldots,k_3$. The regular
representation is given by $\C\Gamma=R_0\oplus R_1\oplus R_2\oplus
R_3$ and thus $k_0+k_1+k_2+k_3$ corresponds to a point in
$\C^3/(\Z_2\times\Z_2)$. 

Computing the value of $w(\gamma)$ from (\ref{eq:a}) for each of the
group elements shows that we expect $H_0$ to be dimension one and
$H_2$ to be dimension three with $H_n$ dimension zero for $n>2$. Thus
the filtration associated to the Atiyah-Hirzebruch spectral sequence
implies that 
\begin{equation}
  H_2 \cong K_0/K^0_0 \cong K_0/H_0.
\end{equation}
That is, we may consider $H_2\cong\Z^3$ to be generated by $k_1,k_2,k_3$ with
a redundant generator $k_0=-k_1-k_2-k_3$.

In order to map this to the classical picture for the homology of this
orbifold, we need to blow-up. The toric picture for a blow-up is
given in figure~\ref{fig:tor10}. We refer to \cite{me:orb2} for
example for more details.

\iffigs
\begin{figure}
  \centerline{\epsfxsize=5cm\epsfbox{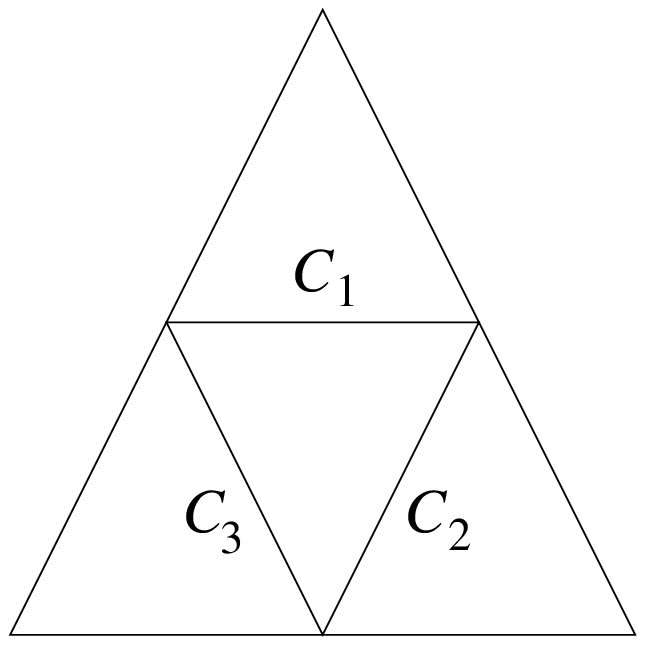}}
  \caption{A blow-up of $\C^3/\Z_2\times\Z_2$.}
  \label{fig:tor10}
\end{figure}
\fi

This resolution contains three isolated rational curves $C_1,C_2,C_3$
shown as lines in figure~\ref{fig:tor10}. We claim that these curves
correspond to the generators $k_1,k_2,k_3$.

This would mean that $C_1$ is represented by the representation
$R_1$. That is, the integers $v_l$ of section \ref{s:quiv} are given by 
$(v_0,v_1,v_2,v_3)=(0,1,0,0)$. Constructing the moduli space for
such a quiver is rather trivial and $\cM_{R_1}$ is a point. This is
consistent with this curve being {\em isolated.}

Now consider the homology class $C_1+C_3$. The geometry of the blow-up
dictates that this is the class of a 2-cycle that is free to move
along the resolution of a fixed {\em line\/} of the orbifold
action. Indeed, for $(v_0,v_1,v_2,v_3)=(0,1,0,1)$ one may show that
$\cM_{R_1\oplus R_3}$ is given by $\C/\Z_2$.

We now show this is consistent with the discussion at the end of
section~\ref{s:mod}. Let $\Gamma_x\cong\Z_2$ be the subgroup of
$\Z_2\times\Z_2$ generated by $a$. This fixes a complex line. Let
$R_x$ be trivial representation of this $\Z_2$ subgroup and let $R_x'$
be the induced representation of $\Gamma$. Now $R_x'=\C^\Gamma
\otimes_{\C^\Gamma_x} R_x$ is easily seen to be two-dimensional and
generated by $1\otimes1$ and $b\otimes1$. With respect to this basis
one computes $1$ and $a$ to be given by the identity matrix; and $b$
and $ab$ is given by $\begin{pmatrix}0&1\\1&0\end{pmatrix}$.

This implies that $R_x'=R_0\oplus R_2$. Similarly if $R_x$ were the
nontrivial representation of $\Z_2$ then $R_x'=R_1\oplus R_3$. Thus
the regular representation can be decomposed $\C\Gamma=(R_0\oplus
R_2)\oplus(R_1\oplus R_3)$ to give two wrapped branes running up and
down the lines fixed by $a$. Note that in homology
$k_0+k_2=-(k_1+k_3)$ and so these two branes are oppositely oriented
consistent with the description in \cite{DDG:wrap}. 

The K-theory story
gives the full picture. The representations $R_0\oplus R_2$ and
$R_1\oplus R_3$ both correspond to the homology element given by the
two-cycle $C_1+C_3$. The sum of these representations gives a point
(rather than a trivial cycle).

In the case of the regular representation we have various
components for $\cM_{R_0\oplus R_1\oplus R_2\oplus R_3}$:
\begin{enumerate}
\item A point may move anywhere in $\C^3/(\Z_2\times\Z_2)$. This gives
a component of the moduli space equal to $\C^3/(\Z_2\times\Z_2)$.
\item Two wrapped branes may move up and down the complex line fixed
by $a$. This gives a component equal to $\C^2/(\Z_2)^3$.
\item Two wrapped branes may move up and down the complex line fixed
by $b$. This gives another component equal to $\C^2/(\Z_2)^3$.
\item Two wrapped branes may move up and down the complex line fixed
by $ab$. This gives another component equal to $\C^2/(\Z_2)^3$.
\end{enumerate}

\subsubsection{With discrete torsion}

Now let us switch discrete torsion on. The discrete quaternion group
$\mathsf{H}$ can be written as a central extension:
\begin{equation}
  1\to\Z_2\to\mathsf{H}\to\Z_2\times\Z_2.
\end{equation}
This defines a cocycle in $H^2(\Z_2\times\Z_2,\Z_2)$ which defines a
cocycle $\alpha\in H^2(\Z_2\times\Z_2,\GU(1))$. The cocycle is given in
table~\ref{tab:d1}.\footnote{It is {\em not\/} possible to 
write a valid cocycle in any way as $\alpha(a^m
b^n,a^{m'}b^{n'})=i^{mn'-m'n}$ as 
has been claimed in the literature.} This case was analyzed in
\cite{D:disctor}. 

\begin{table}
\def\p{\phantom{-}}
$$\lower5mm\hbox{$\gamma_1$}\begin{array}{c|cccc}
\multicolumn{5}{c}{~~~\gamma_2}\\
\multicolumn{1}{c}{}&1&a&b&ab\\
\cline{2-5}
1&1&\p1&\p1&\p1\\
a&1&-1&\p1&-1\\
b&1&-1&-1&\p1\\
ab&1&\p1&-1&-1
\end{array}$$
  \caption{The value of $\alpha(\gamma_1,\gamma_2)$.}
  \label{tab:d1}
\end{table}

The proof of theorem~\ref{th:dt} shows that any irreducible
representation of $\Gamma$ must have a dimension which is a multiple
of two. Since $\Gamma$ has only four elements, (\ref{eq:dim}) implies
that there can only be a
single irreducible representation, $R_0$, of dimension two. The rather
trivial quiver in this case is shown in figure~\ref{fig:q11}.

\iffigs
\begin{figure}
  \centerline{\epsfxsize=3cm\epsfbox{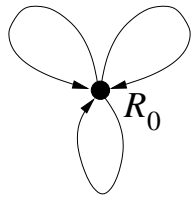}}
  \caption{The quiver for $\Z_2\times\Z_2$ with discrete torsion.}
  \label{fig:q11}
\end{figure}
\fi

The only $\alpha$-regular conjugacy class is the identity.  The value
of $w(1)$ is zero and so we have $H_0$ having rank one.  There are no
twisted strings states which generate $H_n$ for $n>0$. However, it is
not correct to say that $H_0$ is the only nontrivial stringy homology
group.

It is clear that
$\C^\alpha\Gamma=2R_0$. This implies that $K_0/K_0^0\cong\Z_2$ which
means that {\em one of the stringy homology groups $H_{2a}$ is given
by $\Z_2$ for some $a\geq1$.} It is not clear how this torsion cycle
could be declared to be a 2-cycle rather than a 4-cycle etc. All that
we see is that there is a torsion cycle somewhere!

For the wrapped branes, let us consider a subset $\Z_2\subset\Gamma$
which fixes some complex line. Restricting to this subgroup,
$\alpha$-twisted projective representations become linear
representations. Either of the irreducible representations of $\Z_2$
induce the projective representation $R_0$ of $\Gamma$. Thus $R_0$
gives the brane stuck along the fixed lines.

Again in the case of the representation $\C^\alpha\Gamma$ we have various
components for $\cM_{2R_0}$:\footnote{The extra components were missed in
the analysis of \cite{D:disctor}.}
\begin{enumerate}
\item A point may move anywhere in $\C^3/(\Z_2\times\Z_2)$. This gives
a component of the moduli space equal to $\C^3/(\Z_2\times\Z_2)$.
\item Two wrapped branes may move up and down the complex line fixed
by $a$. This gives a component equal to $\C^2/(\Z_2)^3$.
\item Two wrapped branes may move up and down the complex line fixed
by $b$. This gives another component equal to $\C^2/(\Z_2)^3$.
\item Two wrapped branes may move up and down the complex line fixed
by $ab$. This gives another component equal to $\C^2/(\Z_2)^3$.
\item One wrapped brane may move along the line fixed by $a$ and one
wrapped brane may move along the line fixed by $b$. This gives another
component equal to $\C^2/(\Z_2)^3$. 
\item One wrapped brane may move along the line fixed by $a$ and one
wrapped brane may move along the line fixed by $ab$. This gives another
component equal to $\C^2/(\Z_2)^3$. 
\item One wrapped brane may move along the line fixed by $b$ and one
wrapped brane may move along the line fixed by $ab$. This gives another
component equal to $\C^2/(\Z_2)^3$. 
\end{enumerate}
In particular, there are a good deal more components than there were
in the case with no discrete torsion. This is because the two wrapped
branes are not now paired. One may obtain the K-theory element
corresponding to a point by adding {\em any\/} two wrapped branes
together. They need not originate from the same fixed line.

\subsection{A trihedral group}    \label{ss:tri}

Let $a$ and $b$ generate $\Z_2\times\Z_2$ as above. Let
\begin{equation}
  g = \begin{pmatrix}0&1&0\\0&0&1\\1&0&0\end{pmatrix}.
\end{equation}
In this section we will let $\Gamma$ be the twelve element group
generated by $a$, $b$ and $g$. This is a ``trihedral group'' and is
probably the easiest nonabelian to analyze in the context of the McKay
correspondence. It has been analyzed in \cite{Ito:tri,IR:McK,GLR:nonab}
for example.

Note that $g^{-1}ag=b$, $g^{-1}bg=ab$, and $g^{-1}abg=a$. This gives a
$\Z_3$ action on $\Z_2\times\Z_2$ and realizes $\Gamma$ as the
semi-direct product $\Z_3\ltimes(\Z_2\times\Z_2)$. One may now use the
Hochschild-Serre spectral sequence (see section VII.6 of \cite{Brown:}
for example) to show that $H^2(\Gamma,\GU(1))\cong\Z_2$. Thus, as in
the previous example, we have a choice of no discrete torsion or a
unique nontrivial discrete torsion. 

\subsubsection{No discrete torsion}

It is a simple matter to determine that $\Gamma$ has 4 conjugacy classes
given by $\{1\}$, $\{a,b,ab\}$, $\{g,ag,bg,abg\}$ and
$\{g^2,ag^2,bg^2,abg^2\}$. The characters 
are shown in table~\ref{tab:21}, where $\omega=\exp(2\pi i/3)$. $R_0$
is the trivial representation and $R_3$ is the same as $Q$. The quiver
is shown in figure~\ref{fig:q21}.

\begin{table}
\def\p{\phantom{-}}
$$\begin{array}{r|cccc}
\multicolumn{1}{c}{}&R_0&R_1&R_2&R_3\\
\cline{2-5}
1&1&1&1&\p3\\
a,b,ab&1&1&1&-1\\
g,ag,bg,abg&1&\omega&\omega^2&\p0\\
g^2,ag^2,bg^2,abg^2&1&\omega^2&\omega&\p0
\end{array}$$
  \caption{The characters of $\Z_3\ltimes(\Z_2\times\Z_2)$.}
  \label{tab:21}
\end{table}

\iffigs
\begin{figure}
  \centerline{\epsfxsize=5cm\epsfbox{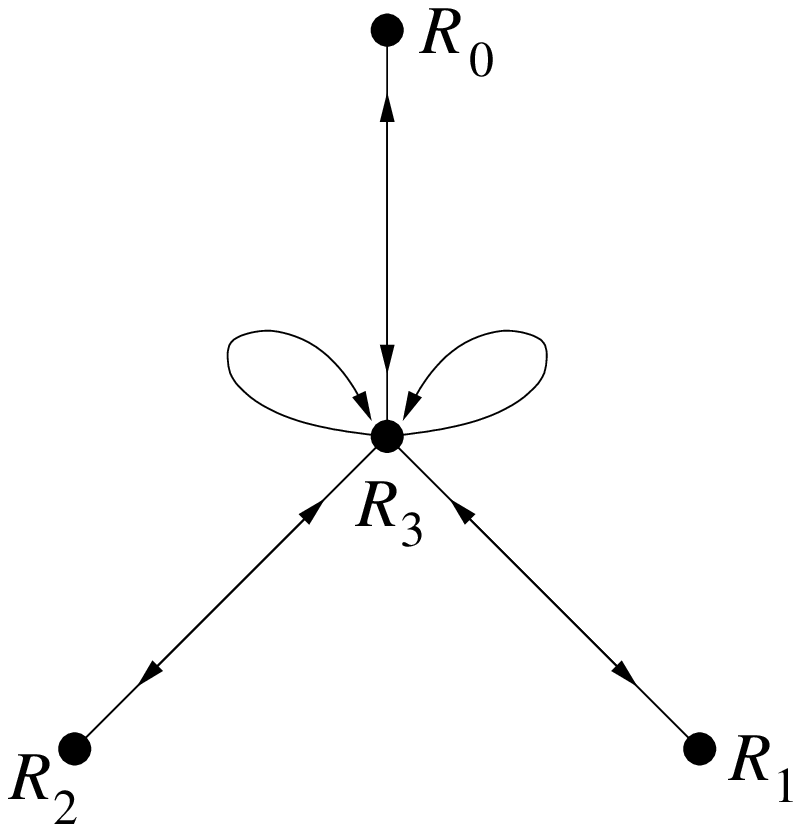}}
  \caption{The quiver for $\Z_3\ltimes(\Z_2\times\Z_2)$ with no
  discrete torsion.} 
  \label{fig:q21}
\end{figure}
\fi

Now $w(1)=0$ and $w(a)=w(g)=w(g^2)=1$. This leads to $H_0\cong\Z$ and
$H_2\cong\Z^3$. 

Next note that the $\Z_3$ subgroup of $\Gamma$ generated by $g$ fixes
a line. Let us find the representations of $\Gamma$ corresponding to
wrapped branes running along this line. A little group theory shows
that the trivial representation of $\Z_3$ induces the representation
$R_0\oplus R_3$ of $\Gamma$ while the other two irreducible
representations of $\Z_3$ induce the representations $R_1\oplus R_3$
and $R_2\oplus R_3$. The representation of a point 
which is $\C\Gamma=R_0\oplus R_1\oplus R_2\oplus3R_3$ may thus break
up into these three wrapped branes in accord with the usual picture.

We also have a line of fixed points generated by the $\Z_2$ subgroup
generated by $a$. (This is identified with the lines fixed by $b$ and
$ab$ by the $\Z_3$ action.) Again, it is an exercise in group theory
to show that the trivial representation of $\Z_2$ induces the
representation $R_0\oplus R_1\oplus R_2\oplus R_3$. The nontrivial
irreducible representation of $\Z_2$ induces the representation
$2R_3$. Again these two representations add up to give the regular
representation. 

Note that for the $\Z_2$-fixed line there is a definite lack of
symmetry between the two wrapped branes. One of them comes from a
representation which is 2-divisible. That is, the state corresponding
to the representation $R_3$ is stuck at the origin where as twice this
representation gives a state that is free to run along the $\Z_2$
fixed line. This is a little reminiscent of torsion even though there
there is no torsion in this model.

\subsubsection{With discrete torsion}

The quaternion group $\mathsf{H}$ admits a $\Z_3$ automorphism leading
to the group $\Z_3\ltimes\mathsf{H}$. This can be written as a central
extension
\begin{equation}
  1\to\Z_2\to\Z_3\ltimes\mathsf{H}\to\Gamma\to1,
	\label{eq:Z3H}
\end{equation}
where $\Gamma$ is our desired trihedral group. This central extension
gives an explicit representation of the group cocycle corresponding to
the nontrivial choice of discrete torsion.

The analysis of section~\ref{ss:K} together with equation
(\ref{eq:Z3H}) tells us that every projective representation of
$\Gamma$ must have an even dimension. Since $\Gamma$ has twelve
elements, (\ref{eq:dim}) implies that there must be three irreducible
projective representations 
each of dimension two. The three $\alpha$-regular conjugacy classes
are $\{1\}$, $\{g,ag,bg,abg\}$ and $\{g^2,ag^2,bg^2,abg^2\}$. The
character table is given in table~\ref{tab:22} and the quiver in
figure~\ref{fig:q22}. 

\begin{table}
$$\begin{array}{r|ccc}
\multicolumn{1}{c}{}&R_0&R_1&R_2\\
\cline{2-4}
1&2&2&2\\
g,ag,bg,abg&-1&-\omega&-\omega^2\\
g^2,ag^2,bg^2,abg^2&-1&-\omega^2&-\omega
\end{array}$$
  \caption{The projective characters of $\Z_3\ltimes(\Z_2\times\Z_2)$.}
  \label{tab:22}
\end{table}

\iffigs
\begin{figure}
  \centerline{\epsfxsize=6cm\epsfbox{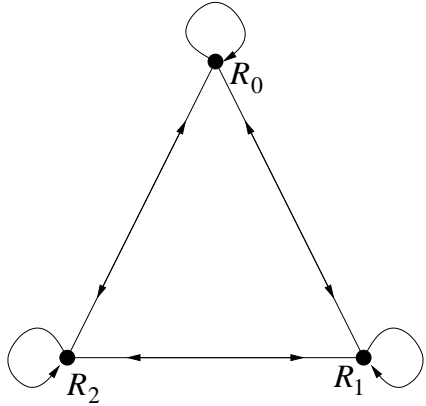}}
  \caption{The quiver for $\Z_3\ltimes(\Z_2\times\Z_2)$ with
  discrete torsion.} 
  \label{fig:q22}
\end{figure}
\fi

The fields twisted by $g$ and $g^2$ both have $w=1$. Thus we appear to
have two generators for $H_2$. The representation $\C^\alpha\Gamma$ is
2-divisible and so we have 2-torsion somewhere in $H_*$. Either one
writes $H_2\cong\Z\oplus\Z\oplus\Z_2$, or the 2-torsion goes into a
higher $H_n$. As in the previous case with discrete torsion, it is
difficult to ascribe any dimensionality to this torsion cycle.

The element $g$ fixes a complex line as before. This $\Z_3$ subgroup's
three irreducible representations induce $R_1\oplus R_2$, $R_0\oplus
R_2$ and $R_0\oplus R_1$ 
respectively as the projective representations of $\Gamma$. These are
our three wrapped branes that move on this complex line.

The element $a$ fixes another complex line as before. The two
irreducible representations of the corresponding $\Z_2$ both induce
the projective representation $R_1\oplus R_2\oplus R_3$. This
representation is ``half of a point'' and gives a state that is free to
move along this fixed line. Twice this state can move anywhere.


\section{Discussion of Strings versus D-Branes}  \label{s:conc}

We would like to draw attention to the fact that theorem~\ref{th:dt}
does {\em not\/} say that the torsion in the homology of an orbifold
with discrete torsion is given by $H^2(\Gamma,\GU(1))$. Rather we say
that $\alpha$ may take on any value of $H^2(\Gamma,\GU(1))$ and, for a
given choice, the homology of the orbifold will naturally contain a
cyclic torsion component $\Z_p$ where $p$ depends upon $\alpha$.

The examples in section~\ref{s:eg} both had
$H^2(\Gamma,\GU(1))\cong\Z_2$ and the nontrivial choice of $\alpha$
yielded $p=2$. Life can get more complicated than this however. It is
quite possible that $H^2(\Gamma,\GU(1))$ is not a cyclic group. For
$\Gamma=\Z_2\times\Z_2\times\Z_2\subset\SU(4)$, one has
$H^2(\Gamma,\GU(1))\cong(\Z_2)^3$ for example.\footnote{It would be
interesting to see if any finite subgroup of $\SU(3)$ gives a
noncyclic discrete torsion group.} The torsion in the homology we
found in section~\ref{s:McKay} is cyclic and can {\em never\/}
therefore equal the 
full discrete torsion group in this case. (We should point out that
there may be other contributions to torsion in the homology groups of
dimension $>2$ but this is not the torsion in homology we are
naturally associating to the discrete torsion.)

This, combined with the K-theory nature of D-brane charge gives a
definite asymmetry between the string and the D1-brane as we now
discuss. 

Consider a type IIB string on a space $Y$.
The $B$-field which is associated to the string charge is
valued in $H^2(Y,\GU(1))$. It generally believed that the $B$ field 
therefore encodes the discrete torsion degree of freedom. This has
been established rigorously by Sharpe \cite{Sh:gerb1,Sh:gerb2,Sh:gerbDT} if one
views the $B$-field as a gerbe connection.

The analogue degree of freedom for the D1-brane comes from the RR
degrees of freedom which live on a torus associated to the lattice of
D-brane charges. In the case of $Y=\C^n/\Gamma$ we have
proposition~\ref{p:McKK} which implies that this torus has dimension
given by the number of irreducible (projective) representations of
$\Gamma$. Two points are worth noting:
\begin{enumerate}
\item This torus is a connected space --- there are never discrete
degrees of freedom. This is because the lattice of charges in
proposition~\ref{p:McKK} is a free group.
\item The dimensionality depends upon a choice $\alpha$ of discrete
torsion.
\end{enumerate}

Clearly this implies that there is no analogue of ``discrete torsion''
in the RR sector. The $B$-field degrees of freedom and the RR degrees
of freedom are completely different. We believe that the
interpretation of this fact is that {\em one cannot
truly claim that there is an S-duality of the type IIB string which
exchanges the string with the D1-brane.} Note that S-duality was also
analyzed in a related context in \cite{DMK:KMbig}. It would be
interesting to understand the relation between their work and ours.

Note that this is consistent with the point of view that such a
duality is likely to be killed if we look at vacua with too little
supersymmetry \cite{AP:T,BGHL:IIBS,me:hyp}. In order to obtain
nontrivial discrete torsion we must compactify on a space with
holonomy at least $\SU(3)$.\footnote{For any finite group
$\Gamma\subset\SU(2)$ one may show that $H^2(\Gamma,\GU(1))$ is
trivial. This may be done by applying the Cartan--Leray spectral
sequence to the free quotient $S^3/\Gamma$.} Thus we only see these
peculiarities in the discrete degrees of freedom for theories with
eight supercharges or fewer.

Note finally that for examples where the orbifold action is free, one
might expect discrete degrees of freedom in the RR sector. An example
of this was analyzed in \cite{FHSV:N=2,me:flower} where are definite
choice of an apparent discrete degree of freedom was required in order
to obtain ``black hole level matching''. Clearly we do not yet
completely understand the RR degrees of freedom in general.


\section*{Acknowledgements}

It is a pleasure to thank R.~Bryant, D.~Hain, E.~Sharpe and everyone
in ``The McKay Correspondence'' workshop at UNC for useful
conversations.  Research partially supported by National Science
Foundation grant DMS-0074072.  
P.S.A.\ is also supported in part by a research fellowship from the Alfred
P.~Sloan Foundation. 


\begin{thebibliography}{10}

\bibitem{BDS:prob}
T.~Banks, M.~R. Douglas, and N.~Seiberg,
\newblock {\em Probing F-theory with Branes},
\newblock Phys. Lett. {\bf B387} (1996) 278--281, hep-th/9605199.

\bibitem{DM:qiv}
M.~R. Douglas and G.~Moore,
\newblock {\em D-branes, Quivers, and ALE Instantons},
\newblock hep-th/9603167.

\bibitem{Doug:Dgeom}
M.~R. Douglas,
\newblock {\em Topics in D-geometry},
\newblock Class. Quant. Grav. {\bf 17} (2000) 1057--1070, hep-th/9910170.

\bibitem{Reid:McK}
M.~Reid,
\newblock {\em La Correspondance de McKay},
\newblock math.AG/9911165,
\newblock S{\'e}minaire Bourbaki, 52{\`e}me ann{\'e}e, novembre 1999, no. 867,
  to appear in Ast{\'e}risque 2000.

\bibitem{DGM:Dorb}
M.~R. Douglas, B.~R. Greene, and D.~R. Morrison,
\newblock {\em Orbifold Resolution by D-Branes},
\newblock Nucl. Phys. {\bf B506} (1997) 84--106, hep-th/9704151.

\bibitem{DG:fracM}
D.-E. Diaconescu and J.~Gomis,
\newblock {\em Fractional Branes and Boundary States in Orbifold Theories},
\newblock hep-th/9906242.

\bibitem{D:disctor}
M.~R. Douglas,
\newblock {\em D-branes and Discrete Torsion},
\newblock hep-th/9807235.

\bibitem{Vafa:tor}
C.~Vafa,
\newblock {\em Modular Invariance and Discrete Torsion on Orbifolds},
\newblock Nucl. Phys. {\bf B273} (1986) 592--606.

\bibitem{Gomis:dt}
J.~Gomis,
\newblock {\em D-branes on Orbifolds with Discrete Torsion and Topological
  Obstruction},
\newblock JHEP {\bf 05} (2000) 006, hep-th/0001200.

\bibitem{Joyce:desing}
D.~Joyce,
\newblock {\em On the Topology of Desingularizations of Calabi--Yau Orbifolds},
\newblock math.AG/ 9806146.

\bibitem{BL:}
D.~Berenstein and R.~G. Leigh,
\newblock {\em Discrete torsion, AdS/CFT and Duality},
\newblock J. High Energy Phys. {\bf 01} (2000) 038, hep-th/0001055.

\bibitem{Ruan:dt}
Y.~Ruan,
\newblock {\em Discrete torsion and twisted orbifold cohomology},
\newblock math.AG/0005299.

\bibitem{SI:I}
A.~V. Sardo~Infirri,
\newblock {\em Partial Resolutions of Orbifold Singularities via Moduli Spaces
  of HYM-type Bundles},
\newblock alg-geom/9610004.

\bibitem{SI:II}
A.~V. Sardo~Infirri,
\newblock {\em Resolutions of Orbifold Singularities and the Transportation
  Problem on the McKay Quiver},
\newblock alg-geom/9610005.

\bibitem{DDG:wrap}
D.-E. Diaconescu, M.~R. Douglas, and J.~Gomis,
\newblock {\em Fractional Branes and Wrapped Branes},
\newblock JHEP {\bf 02} (1998) 013, hep-th/9712230.

\bibitem{DF:dt2}
M.~R. Douglas and B.~Fiol,
\newblock {\em D-branes and Discrete Torsion. II},
\newblock hep-th/9903031.

\bibitem{BJL:dtdef}
D.~Berenstein, V.~Jejjala, and R.~G. Leigh,
\newblock {\em Marginal and Relevant Deformations of N = 4 Field Theories},
\newblock hep-th/0005087.

\bibitem{AMG:stab}
P.~S. Aspinwall and D.~R. Morrison,
\newblock {\em Stable Singularities in String Theory},
\newblock Commun. Math. Phys. {\bf 178} (1996) 115--134, hep-th/9503208,
\newblock appendix by M. Gross.

\bibitem{Karpil:proj}
G.~Karpilovsky,
\newblock {\em Projective Representations of Finite Groups},
\newblock Dekker, 1985.

\bibitem{GLR:nonab}
B.~R. Greene, C.~I. Lazaroiu, and M.~Raugas,
\newblock {\em D-branes on Nonabelian Threefold Quotient Singularities},
\newblock Nucl. Phys. {\bf B553} (1999) 711--749, hep-th/9811201.

\bibitem{VW:tor}
C.~Vafa and E.~Witten,
\newblock {\em On Orbifolds with Discrete Torsion},
\newblock J. Geom. Phys. {\bf 15} (1995) 189--214, hep-th/9409188.

\bibitem{W:phase}
E.~Witten,
\newblock {\em Phases of $N=2$ Theories in Two Dimensions},
\newblock Nucl. Phys. {\bf B403} (1993) 159--222, hep-th/9301042.

\bibitem{AGM:II}
P.~S. Aspinwall, B.~R. Greene, and D.~R. Morrison,
\newblock {\em \CY\ Moduli Space, Mirror Manifolds and Spacetime Topology
  Change in String Theory},
\newblock Nucl. Phys. {\bf B416} (1994) 414--480.

\bibitem{AGM:sd}
P.~S. Aspinwall, B.~R. Greene, and D.~R. Morrison,
\newblock {\em Measuring Small Distances in $N=2$ Sigma Models},
\newblock Nucl. Phys. {\bf B420} (1994) 184--242, hep-th/9311042.

\bibitem{BKM:MisM}
T.~Bridgeland, A.~King, and M.~Reid,
\newblock {\em Mukai implies McKay},
\newblock math.AG/9908027.

\bibitem{MM:K}
R.~Minasian and G.~Moore,
\newblock {\em K-Theory and Ramond-Ramond Charge},
\newblock J. High Energy Phys. {\bf 11} (1997) 002, hep-th/9710230.

\bibitem{W:K}
E.~Witten,
\newblock {\em D-branes and K-Theory},
\newblock J. High Energy Phys. {\bf 12} (1998) 019, hep-th/9810188.

\bibitem{Hatch:top}
A.~Hatcher,
\newblock {\em Algebraic Topology I},
\newblock Cambridge, 2000,
\newblock to appear, currently available at {\tt
  http://www.math.cornell.edu/\~{ }hatcher/\#ATI}.

\bibitem{AH:K}
M.~F. Atiyah and F.~Hirzebruch,
\newblock {\em Vector Bundles and Homogeneous Spaces},
\newblock Proc. Sympos. Pure Math. {\bf III} (1961) 7--38.

\bibitem{DMK:KMbig}
D.-E. Diaconescu, G.~Moore, and E.~Witten,
\newblock {\em $E_8$ Gauge Theory, and a Derivation of K-Theory from M-Theory},
\newblock hep-th/0005090.

\bibitem{Cald:dK}
A.~Caldararu,
\newblock {\em Derived Categories of Twisted Sheaves on Calabi--Yau Manifolds},
\newblock PhD thesis, Cornell, 2000.

\bibitem{BRG:Z2}
B.~R. Greene,
\newblock {\em D-brane Topology Changing Transitions},
\newblock Nucl. Phys. {\bf B525} (1998) 284--296, hep-th/9711124.

\bibitem{me:orb2}
P.~S. Aspinwall,
\newblock {\em Resolution of Orbifold Singularities in String Theory},
\newblock in B.~Greene and S.-T. Yau, editors, ``Mirror Symmetry II'', pages
  355--426, International Press, 1996,
\newblock hep-th/9403123.

\bibitem{Ito:tri}
Y.~Ito,
\newblock {\em Crepent Resolutions of Trihedral Singularities},
\newblock Proc. Japan Acad. {\bf 70} (1994) 131--136.

\bibitem{IR:McK}
Y.~Ito and M.~Reid,
\newblock {\em The McKay Correspondence for Finite Subgroups of $\Sl(3,\C)$},
\newblock in M.~Andreatta et~al., editors, ``Higher Dimensional Complex
  Varieties'', pages 221--240, de Gruyter, 1996,
\newblock alg-geom/9411010.

\bibitem{Brown:}
K.~S. Brown,
\newblock {\em Cohomology of Groups}, Graduate Texts in Mathematics~{\bf 87},
\newblock Springer, 1982.

\bibitem{Sh:gerb1}
E.~R. Sharpe,
\newblock {\em Discrete Torsion and Gerbes. I},
\newblock hep-th/9909108.

\bibitem{Sh:gerb2}
E.~R. Sharpe,
\newblock {\em Discrete Torsion and Gerbes. II},
\newblock hep-th/9909120.

\bibitem{Sh:gerbDT}
E.~R. Sharpe,
\newblock {\em Discrete Torsion},
\newblock hep-th/0008154.

\bibitem{AP:T}
P.~S. Aspinwall and M.~R. Plesser,
\newblock {\em T-Duality Can Fail},
\newblock J. High Energy Phys. {\bf 08} (1999) 001, hep-th/9905036.

\bibitem{BGHL:IIBS}
R.~Boehm, H.~Guenther, C.~Herrmann, and J.~Louis,
\newblock {\em Compactification of Type IIB String Theory on Calabi--Yau
  Threefolds},
\newblock hep-th/9908007.

\bibitem{me:hyp}
P.~S. Aspinwall,
\newblock {\em Aspects of the Hypermultiplet Moduli Space in String Duality},
\newblock J. High Energy Phys. {\bf 04} (1998) 019, hep-th/9802194.

\bibitem{FHSV:N=2}
S.~Ferrara, J.~Harvey, A.~Strominger, and C.~Vafa,
\newblock {\em Second Quantized Mirror Symmetry},
\newblock Phys. Lett. {\bf 361B} (1995) 59--65, hep-th/9505162.

\bibitem{me:flower}
P.~S. Aspinwall,
\newblock {\em An $N=2$ Dual Pair and a Phase Transition},
\newblock Nucl. Phys. {\bf B460} (1996) 57--76, hep-th/9510142.

\end{thebibliography}

\end{document}